\documentclass[12pt]{article}
\usepackage[english]{babel}
\usepackage{amsmath}
\usepackage{amssymb}
\usepackage{bbm}
\usepackage{color}
\usepackage[titletoc,toc,title]{appendix}
\Roman{section}
\newcommand{\naw}[1]{\left(#1\right)}

\newcommand{\com}[1]{\left[#1\right]}

\newcommand{\poisson}[1]{\left\{#1\right\}}

\begin{document}

\begin{center}
\textsc{\Large{On the quantization of nonlocal theory}}
\newline

\large{Katarzyna Bolonek-Laso\'n}\footnote{kbolonek1@wp.pl}\\ 
\emph{\normalsize{Faculty of Economics and Sociology, Department of Statistical Methods,\\ University of Lodz, Poland.}}\\
\large{Piotr Kosi\'nski}\footnote{pkosinsk@uni.lodz.pl}\\
\emph{\normalsize{Faculty of Physics and Applied Informatics,\\ Department of Informatics,\\ University of Lodz, Poland.}}
\end{center}

\begin{abstract}
A simple nonlocal theory is put into Hamiltonian form and quantized by using the modern version of Ostrogradski approach.
\end{abstract}

\section{Introduction}
Some attention has been attracted in the last decade by the problem of quantizing the models defined by the Lagrangians which are nonlocal in time. This is mainly due to the fact that such nonlocalities do appear in field theories on noncommutative space-time \cite{Douglas}. However, the very problem dates back to the seminal paper of Pais and Uhlenbeck \cite{Pais}. They have shown how the quadratic nonlocal Lagrangians can be converted into (alternating) sum of independent oscillators (possibly complex). The corresponding transformation is defined slightly formally in terms of infinite order differential operators which single out particular frequencies of classical motion. Once the Lagrangian is represented as a sum of independent oscillators, the Hamiltonian form of dynamics can be readily constructed and quantization performed. However, the method seem to be applicable only to some particular models.

The Hamiltonian formalism for general higher-derivative theories has been constructed long time ago by Ostrogradski \cite{Ostrogradski}. Its more sophisticated form, in principle applicable to arbitrary nonlocal theories, has been given by a number of authors \cite{Llosa}$\div$\cite{Woodard}. Its applicability is restricted by the fact that the original dynamical equations appear here as a constraint which has to be dealt with by use of Dirac formalism. This is very difficult because the explicit solutions to the equations of motion are hardly known. However, for particular linear theories one is able to find solutions which allows to describe explicitly the dynamics in Hamiltonian form.

It has been noted quite recently that the Pais-Uhlenbeck model exhibits a remarkable symmetry provided its eigenfrequencies are proportional to the consecutive odd integers. More precisely, the Noether symmetries of the Pais-Uhlenbeck oscillator of order $2l+1$ ($l$ being half integer) form the so called $l$-conformal Galilei or Newton-Hooke group \cite{Gala}$\div$\cite{Andrzejewski2} which gained recently much attention. The nonlocal model introduced by Pais and Uhlenbeck provides a natural infinite order generalization of their finite order oscillators with special eigenfrequencies. Therefore, it deserves a more careful study. In particular, it is interesting to quantize it using the generalized Ostrogradski formalism which would also provide proper framework for studying the relevant symmetries. 

In the present note we use the method developed in Refs. \cite{Llosa}$\div$\cite{Woodard} to put into Hamiltonian form and quantize the nonlocal model Pais and Uhlenbeck. The final results coincide, as expected, with those of Pais and Uhlenbeck  but the method used is quite different. The model is so simple that all details of the generalized Ostrogradski method can be revealed.

\section{Hamiltonian formalism and quantization}
Our starting point is the following nonlocal lagrangian
\begin{equation}
L=-\frac{m}{\alpha^2}q\naw{t}q\naw{t+\alpha}\label{a}
\end{equation}
where $m$ and $\alpha$ are some constants of dimensions of mass and time, respectively. $L$ provides a natural generalization of standard harmonic oscillator. Indeed, expanding $q\naw{t+\alpha}$ to the second order in $\alpha$ one obtains
\begin{equation}
L=\frac{m\dot{q}^2}{2}-\frac{m\omega^2q^2}{2}+\frac{d}{dt}\naw{\frac{m}{2\alpha}q^2-\frac{m}{2}q\dot{q}}+O\naw{\alpha}.
\end{equation}
Skipping total derivative one obtains harmonic oscillator with the frequency $\omega^2\equiv\frac{2}{\alpha^2}$.\\
The equation of motion
\begin{equation}
\delta S\equiv\int_{-\infty}^{\infty}dt'\frac{\delta L\naw{t'}}{\delta q\naw{t}}=0
\end{equation}
following from eq. (\ref{a}) reads
\begin{equation}
q\naw{t-\alpha}+q\naw{t+\alpha}=0.
\end{equation}   
In order to quantize our theory one has to put it into Hamiltonian form. To this end we use the formalism proposed in Refs. \cite{Llosa}$\div$\cite{Gomis1} which provides a far-reaching generalization of the Ostrogradski approach \cite{Ostrogradski}. According to the prescription of Refs. \cite{Llosa}$\div$\cite{Gomis1} one introduces a continuous index $\lambda$ and makes the following replacements
\begin{equation}
\begin{split}
& q\naw{t}\rightarrow Q\naw{t,\lambda},\qquad q\naw{t+\alpha}\rightarrow Q\naw{t,\lambda+\alpha}\\
& \dot{q}\naw{t}\rightarrow Q'\naw{t,\lambda}\equiv\frac{\partial Q\naw{t,\lambda}}{\partial\lambda},\qquad\dot{q}\naw{t+\alpha}\rightarrow Q'\naw{t,\lambda+\alpha}.
\end{split}\label{a1}
\end{equation}
The lagrangian is defined as
\begin{equation}
\widetilde{L}\naw{t}\equiv\int_{-\infty}^{\infty}d\lambda\delta\naw{\lambda}L\naw{Q\naw{t,\lambda},Q'\naw{t,\lambda},...}
\end{equation}
where $L\naw{Q\naw{t,\lambda},Q'\naw{t,\lambda},...}$ is obtained from the original Lagrangian by making the replacements (\ref{a1}).

In our case
\begin{equation}
L\naw{Q\naw{t,\lambda},Q'\naw{t,\lambda},...}=-\frac{m}{\alpha^2}Q\naw{t,\lambda}Q\naw{t,\lambda+\alpha}
\end{equation}
and
\begin{equation}
\widetilde{L}\naw{t}=-\frac{m}{\alpha^2}Q\naw{t,0}Q\naw{t,\alpha}.
\end{equation}
The Hamiltonian and the Poisson bracket read
\begin{equation}
\begin{split}
&H\naw{t}\equiv\int_{-\infty}^{\infty}d\lambda P\naw{t,\lambda}Q'\naw{t,\lambda}-\widetilde{L}\naw{t}=\\
& =\int_{-\infty}^{\infty}d\lambda P\naw{t,\lambda}Q'\naw{t,\lambda}+\frac{m}{\alpha^2}Q\naw{t,0}Q\naw{t,\alpha}
\end{split}\label{a2}
\end{equation}
\begin{equation}
\poisson{Q\naw{t,\lambda},P\naw{t,\lambda'}}=\delta\naw{\lambda-\lambda'}.\label{a3}
\end{equation}
Eqs. (\ref{a2}) and (\ref{a3}) generate the following dynamics
\begin{equation}
\dot{Q}\naw{t,\lambda}=Q'\naw{t,\lambda}\label{b2}
\end{equation}
\begin{equation}
\dot{P}\naw{t,\lambda}=P'\naw{t,\lambda}-\frac{m}{\alpha^2}\naw{\delta\naw{\lambda}Q\naw{t,\alpha}+\delta\naw{\lambda-\alpha}Q\naw{t,0}}.
\end{equation}
In order to recover the original dynamics one has to impose new constraints. First, we define the primary momentum constraints
\begin{equation}
P\naw{t,\lambda}-\frac{1}{2}\int_{-\infty}^{\infty}d\sigma\naw{\text{sgn}\naw{\lambda}-\text{sgn}\naw{\sigma}}\varepsilon\naw{t,\sigma,\lambda}\approx 0\label{a4}
\end{equation}
where 
\begin{equation}
\varepsilon\naw{t,\sigma,\lambda}\equiv\frac{\delta L\naw{Q\naw{t,\sigma},Q'\naw{t,\sigma}...}}{\delta Q\naw{t,\lambda}}.\label{a5}
\end{equation}
In our case eqs. (\ref{a4}) and (\ref{a5}) imply
\begin{equation}
P\naw{t,\lambda}-\frac{m}{2\alpha^2}\naw{\text{sgn}\naw{\lambda-\alpha}-\text{sgn}\naw{\lambda}}Q\naw{t,\lambda-\alpha}\approx 0.\label{aa}
\end{equation}
By differentiating a number of times with respect to time one obtains the secondary constraints
\begin{equation}
\int_{-\infty}^{\infty}d\sigma\varepsilon\naw{t,\sigma,\lambda}=0
\end{equation}
giving here
\begin{equation}
Q\naw{t,\lambda-\alpha}+Q\naw{t,\lambda+\alpha}=0.\label{aa1}
\end{equation}
Summarizing, we have the following set of constraints
\begin{equation}
\varphi_1\naw{t,\lambda}\equiv P\naw{t,\lambda}-\frac{m}{2\alpha^2}\naw{\text{sgn}\naw{\lambda-\alpha}-\text{sgn}\naw{\lambda}}Q\naw{t,\lambda-\alpha}\approx 0
\end{equation}
\begin{equation}
\varphi_2\naw{t,\lambda}\equiv Q\naw{t,\lambda-\alpha}+Q\naw{t,\lambda+\alpha}\approx 0.\label{c}
\end{equation}
In order to convert the constraints, which are here of second kind, into strong in equalities one can define the Dirac bracket
\begin{equation}
\poisson{A,B}_D\equiv\poisson{A,B}-\int_{-\infty}^{\infty}\int_{-\infty}^{\infty}d\lambda d\lambda'\poisson{A,\varphi_i\naw{\lambda}}C_{ij}^{-1}\naw{\lambda,\lambda'}\poisson{\varphi_j\naw{\lambda'},B}\label{b}
\end{equation} 
where
\begin{equation}
C\naw{\lambda,\lambda'}=\left[\begin{array}{cc}
\poisson{\varphi_1\naw{\lambda},\varphi_1\naw{\lambda'}} & \poisson{\varphi_1\naw{\lambda},\varphi_2\naw{\lambda'}}\\
\poisson{\varphi_2\naw{\lambda},\varphi_1\naw{\lambda'}} & \poisson{\varphi_2\naw{\lambda},\varphi_2\naw{\lambda'}}\end{array}\right]
\end{equation}
and
\begin{equation}
\int_{-\infty}^{\infty}d\lambda''C_{ij}\naw{\lambda,\lambda''}C_{jk}^{-1}\naw{\lambda'',\lambda'}=\delta_{ik}\delta\naw{\lambda-\lambda'}.
\end{equation}
However, instead of proceeding via direct computation as sketched above one can do better. First $\varphi_1$ can be used to solve explicitly for $P\naw{t,\lambda}$. Then the reduced phase space is spanned by $Q\naw{t,\lambda}$ subject to the constraint $\varphi_2$. The Hamiltonian (\ref{a2}), expressed in terms of basic coordinates, reads
\begin{equation}
H\naw{t}=\int_{-\infty}^{\infty}d\lambda\Delta\naw{\lambda,\alpha}Q\naw{t,\lambda-\alpha}Q'\naw{t,\lambda}+\frac{m}{\alpha^2}Q\naw{t,0}Q\naw{t,\alpha}\label{b1}
\end{equation}
where
\begin{equation}
\Delta\naw{\lambda,\alpha}\equiv\frac{m}{2\alpha^2}\naw{\text{sgn}\naw{\lambda-\alpha}-\text{sgn}\naw{\lambda}}.
\end{equation}
Due to the constraints $\varphi_2$ the original dynamics is recovered provided the Hamiltonian equations implied by (\ref{b}) and (\ref{b1}) coincide with eq. (\ref{b2}). This is the first condition imposed on Dirac bracket
\begin{equation}
\poisson{Q\naw{t,\lambda},Q\naw{t,\lambda'}}=F\naw{\lambda,\lambda'}.
\end{equation} 
The remaining ones are
\begin{equation}
F\naw{\lambda,\lambda'}=-F\naw{\lambda',\lambda}\label{b3}
\end{equation}
\begin{equation}
F\naw{\lambda-\alpha,\lambda'}+F\naw{\lambda+\alpha,\lambda'}=0,\label{b4}
\end{equation}
the latter being the consequence of the strong equality $\varphi_2=0$.\\
The unique solution to eqs. (\ref{b3}), (\ref{b4}) which produces eq. (\ref{b2}) reads
\begin{equation}
F\naw{\lambda,\lambda'}=\frac{\alpha^2}{m}\sum_{k=-\infty}^{\infty}\naw{-1}^k\delta\naw{\lambda-\lambda'+\naw{2k+1}\alpha}.
\end{equation}
The remaining brackets are easily recovered by demanding that they respect constraints
\begin{equation}
\poisson{Q\naw{t,\lambda},P\naw{t,\lambda'}}_D=\Delta\naw{\lambda',\alpha}F\naw{\lambda,\lambda'-\alpha}
\end{equation}
\begin{equation}
\poisson{P\naw{t,\lambda},P\naw{t,\lambda'}}_D=\Delta\naw{\lambda,\alpha}\Delta\naw{\lambda',\alpha}F\naw{\lambda,\lambda'}.
\end{equation}
We shall now solve explicitly the second constraint. To this end we define new variables
\begin{equation}
\widetilde{Q}\naw{t,\lambda}\equiv Q\naw{t,\lambda}e^{-\frac{i\pi}{2\alpha}\lambda}.
\end{equation}
Then, by virtue of eq. (\ref{c}) $\widetilde{Q}\naw{t,\lambda}$ is periodic in $\lambda$, the period being $2\alpha$. It can be expanded in Fourier series. Taking this into account we conclude that $Q\naw{t,\lambda}$ can be expanded as follows 
\begin{equation}
Q\naw{t,\lambda}=\sum_{n=-\infty}^{\infty}a_n\naw{t}\Psi_n\naw{\lambda}\label{c2}
\end{equation}
where
\begin{equation}
\Psi_n\naw{\lambda}\equiv\frac{1}{\sqrt{2\alpha}}e^{\frac{i\pi}{2\alpha}\naw{2n+1}\lambda}
\end{equation}
$\poisson{\Psi_n\naw{\lambda}}_{n=-\infty}^{\infty}$ form an orthonormal
\begin{equation}
\int_{-\alpha}^{\alpha}d\lambda\overline{\Psi}_n\naw{\lambda}\Psi_m\naw{\lambda}=\delta_{nm}\label{c3}
\end{equation}
and complete set. Additionally,
\begin{equation}
\overline{\Psi}_n\naw{\lambda}=\Psi_{-\naw{n+1}}\naw{\lambda}\label{c1}
\end{equation}
By virtue of eq. (\ref{c1}) the reality condition for $Q\naw{t,\lambda}$ reads
\begin{equation}
\overline{a}_n=a_{-\naw{n+1}}
\end{equation}
$a_n's$ provide new dynamical variables. Eqs. (\ref{c2}) and (\ref{c3}) imply
\begin{equation}
a_n\naw{t}=\int_{-\alpha}^{\alpha}d\lambda\overline{\Psi}_m\naw{\lambda}Q\naw{t,\lambda}.
\end{equation}
It is now straightforward to compute the Dirac bracket for new variables
\begin{equation}
\poisson{a_m,a_n}=\frac{i\alpha^2}{m}\naw{-1}^m\delta_{m+n+1,0}.
\end{equation}
Also the Hamiltonian is easily expressible in terms of new variables
\begin{equation}
H=\frac{m}{2\alpha^2}\sum_{k=-\infty}^{\infty}\naw{-1}^k\frac{\pi}{2\alpha}\naw{2k+1}a_ka_{-\naw{k+1}}.
\end{equation}
By defining
\begin{equation}
a_n=\left\{ \begin{array}{cc}
\frac{\alpha}{\sqrt{m}}c_n, & n-odd\quad positive\\
\frac{\alpha}{\sqrt{m}}c_{-\naw{n+1}}, & n-odd\quad negative\\
\frac{\alpha}{\sqrt{m}}\overline{c}_n, & n-even\quad nonnegative\\
\frac{\alpha}{\sqrt{m}}\overline{c}_{-\naw{n+1}}, & n-even\quad negative\end{array}\right.
\end{equation}
one obtains
\begin{equation}
\poisson{c_m,\overline{c}_n}=-i\delta_{mn}\label{d}
\end{equation}
and
\begin{equation}
H=\sum_{k=0}^{\infty}\naw{-1}^k\frac{\pi}{2\alpha}\naw{2k+1}\overline{c}_kc_k.\label{d1}
\end{equation}
Eqs. (\ref{d}), (\ref{d1}) tell us that our theory is an alternating sum of independent harmonic oscillators of frequencies $\omega_k=\frac{\pi}{2\alpha}\naw{2k+1}$.\\
Eqs. (\ref{d}), (\ref{d1}) can readily quantized. Upon rescaling $c_m\rightarrow\sqrt{\hbar}\hat{c}_m$ and keeping the order of factors as in eq. (\ref{d1}) (which implies the subtraction of zero-energy value) one arrives at 
\begin{equation}
\com{c_m,c_n^+}=\delta_{mn}
\end{equation}
\begin{equation}
H=\hbar\sum_{k=0}^{\infty}\naw{-1}^k\frac{\pi}{2\alpha}\naw{2k+1}c_k^+c_k.
\end{equation}
Our results coincide with those obtained by Pais-Uhlenbeck method.

\section{$\alpha$-expansion}
The model defined by eq. (\ref{a}) can be approximated by finite-order oscillators. To this end let us expand the right hand side of eq. (\ref{a}) in Taylor series in $\alpha$
\begin{equation}
L=-\frac{m^2}{\alpha^2}\sum_{n=0}^{\infty}\frac{\alpha^n}{n!}q\naw{t}q^{\naw{n}}\naw{t}
\end{equation} 
Due to the identity
\begin{equation}
qq^{\naw{2k+1}}=\frac{d}{dt}\naw{\sum_{i=0}^{k-1}q^{\naw{i}}q^{\naw{2k-i}}+\frac{\naw{-1}^k}{2}{q^{\naw{k}}}^2}.
\end{equation}
 $L$ can be rewritten as 
\begin{equation}
L=-\frac{m^2}{\alpha^2}\sum_{n=0}^{\infty}\frac{\alpha^{2n}}{\naw{2n}!}q\naw{t}q^{\naw{2n}}\naw{t}\label{d2}
\end{equation}
or
\begin{equation}
L=-\frac{m^2}{\alpha^2}q\naw{t}\cos\naw{i\alpha\frac{d}{dt}}q\naw{t}.\label{e}
\end{equation}
Consider the Lagrangian $L_k$ obtained by truncating the expansion (\ref{d2}) on k-th term. Let us denote
\begin{equation}
F_k\naw{x}\equiv\sum_{n=0}^{k}\frac{\naw{-1}^nx^{2n}}{\naw{2n}!}.
\end{equation} 
Using $\frac{d^2F_k}{dx^2}=-F_{k-1}$ we find easily that $F_k$ has $2k$ real roots $\pm\nu_n^{\naw{k}}$, $n=1,...,k$. Consequently,
\begin{equation}
L_k=\frac{\naw{-1}^{k-1}m^2\alpha^{2\naw{k-1}}q\naw{t}}{\naw{2k}!}\prod_{n=1}^{k}\naw{\frac{d^2}{dt^2}+\frac{{\nu_n^{\naw{k}}}^2}{\alpha^2}}q\naw{t}.
\end{equation} 
According to Pais and Uhlenbeck the Hamiltonian corresponding to the above Lagrangian can be written as an alternating sum of harmonic oscillators with frequencies $\frac{\nu_{n}^{\naw{k}}}{\alpha}$. By virtue of eq. (\ref{e}):
\begin{equation}
\lim_{k\rightarrow\infty}\nu_n^{\naw{k}}=\naw{2n+1}\frac{\pi}{2}.
\end{equation}
we conclude that taking the limit $k\rightarrow \infty$ in the theories described by the Lagrangians $L_k$ we arrive at the original nonlocal model (\ref{d1}).

\section{Final remarks}
We quantized simple nonlocal model of Pais and Uhlenbeck using the generalized Ostrogradski method described in Refs. \cite{Llosa}$\div$\cite{Woodard}. The main point of this method is that the proper equations of motion are imposed as a constraint (eqs. (\ref{aa}) and (\ref{aa1}) in our case). The reason for the necessity of imposing such a constraint is the following. In the standard Ostrogradski approach, applicable to the theories of arbitrary but finite order, all canonical equations but one, related to the highest time derivative, serve to define the consecutive time derivatives of the initial coordinate variable. The proper equation of motion results from the last Hamiltonian equation related to the highest time derivative. But for systems of infinite order this equation is lacking. This is why one has to introduce the constraint. 

The formalism described above is well suited for the description of symmetries of nonlocal generalization of the set of harmonic oscillators. This will be the subject of the forthcoming paper.


\begin{thebibliography}{99}
\bibitem{Douglas} M.R. Douglas, N. A. Nekrasov,  \emph{Rev. Mod. Phys.} \textbf{73} (2001), 97\\
R J. Szabo, \emph{Phys. Rep.} \textbf{378} (2003), 207
\bibitem{Pais} A. Pais, G. E. Uhlenbeck, \emph{Phys. Rev.}  \textbf{79} (1950), 145
\bibitem{Ostrogradski} M. Ostrogradski, \emph{Mem. Ac. St. Petersburg} \textbf{4} (1850), 385
\bibitem{Llosa} J. Llosa, J. Vives, \emph{J. Math. Phys.} \textbf{35} (1994), 2856
\bibitem{Gomis} J. Gomis, K. Kamimura, J. Llosa, \emph{Phys. Rev.} \textbf{D63} (2001), 045003
\bibitem{Gomis1} J. Gomis, K. Kamimura, T. Ramirez, \emph{Nucl. Phys.} \textbf{B696} (2004), 263
\bibitem{Woodard} R. J. Woodard, \emph{Phys. Rev.} \textbf{A62} (2000), 052105
\bibitem{Gala} A. Galajinsky, I. Masterov, \emph{Nucl. Phys.} \textbf{B866} (2013), 212
\bibitem{Gala1}  A. Galajinsky, I. Masterov, \emph{Phys. Lett.} \textbf{B723} (2013), 190
\bibitem{Andrzejewski} K. Andrzejewski, A. Galajinsky, J. Gonera, I. Masterov, \emph{Nucl. Phys.} \textbf{B885} (2014), 150
\bibitem{Masterov} I. Masterov, \emph{Journ. Math. Phys.} \textbf{55} (2014), 102901
\bibitem{Masterov1} I. Masterov, \emph{Higher-derivative mechanics with N=2 l-conformal supersymmetry}, arXiv:1410.5335
\bibitem{Andrzejewski1} K. Andrzejewski, \emph{Phys. Lett.} \textbf{B738} (2014), 405
\bibitem{Andrzejewski2} K. Andrzejewski, \emph{Nucl. Phys.} \textbf{B889} (2014), 333

\end{thebibliography}
\end{document}